\documentclass[11pt]{article}
\usepackage{graphicx}
\usepackage{type1cm}
\begin{document}
\begin{flushleft}
{\Large
Measurement of airborne radioactivity from the Fukushima reactor accident 
in Tokushima, Japan
}
\\
\vspace{0.5cm}
K.Fushimi, S.Nakayama, M.Sakama$^{2}$ and Y.Sakaguchi$^{3}$
\end{flushleft}
\vspace{0.5cm}
\begin{flushleft}
Institute of Socio Arts and Sciences , The University of Tokushima,
1-1 Minami Josanjimacho Tokushima city, 770-8502 Tokushima, JAPAN
\\
$^{2}$ Department of Radiological Science, Division of Biomedical Sciences, 
Institute of Health Biosciences, The University of Tokushima, 
3-18-15 Kuramotocho Tokushima city, 770-8509 Tokushima, JAPAN
\\
$^{3}$ Faculty of Integrated Arts and Sciences, The University of Tokushima,
1-1 Minami Josanjimacho Tokushima city, 770-8502 Tokushima, JAPAN
\end{flushleft}
\begin{abstract}
The airborne radioactive isotopes from the Fukushima Daiichi nuclear plant was 
measured in Tokushima, western Japan.
The continuous monitoring has been carried out in Tokushima.
From March 23, 2011 the fission product $^{131}$I was observed.
The radioisotopes $^{134}$Cs and $^{137}$Cs were also observed in the 
beginning of April.
However the densities were extremely smaller than the Japanese 
regulation of radioisotopes.
\end{abstract}

\section{Introduction}
Serious damage to the Fukushima Daiichi nuclear power plant 
(141$^{\circ}$27'E, 37$^{\circ}$45'N)
has been caused by huge tsunami followed by the huge earthquake on 11 March 2011.
The plants 1,2 and 3 were operating and the plant 4 was stopped before the earthquake.
The plants made emergency stop just after the earthquake, however, 
all the power plants in Fukushima Daiichi were seriously damaged by the following 
big tsunami.
All the electric power got fault and the cooling system was collapsed.
From 12 March 2012, a large amount of radioactive materials was vented to
avoid more serious damages.
Total amount of vented radioactive isotopes were estimated as $1.5\times 10^{17}$Bq 
for $^{131}$I and $1.2\times 10^{16}$Bq for $^{137}$Cs \cite{TEPCO}.

The largest ejection of radioactivity from the plants occurred on 15th March
and the amount of ejected radioactivity decreased after 17 March\cite{TEPCO}.
During 15 and 16 March, the wind direction changed from north and south, 
the wind direction raised the serious pollution in Iidate village and north Kanto 
district.
After 16 March, the wind direction changed to west and continued for a few days.
The westly wind brought the radioactive isotopes to the Northern Hemisphere.

In the present paper, the measurement of airborne radioactive isotopes in Tokushima which is placed in
western district of Japan is reported.
The arrival date of radioactivity in the world was analyzed to investigate the 
behavior of plume exhausted from the reactor.
The precise information for the detection efficiency of gamma ray was determined 
to analyze the radioactive isotopes.
The detection efficiency of gamma rays were precisely estimated by Monte Carlo 
simulation.
The coincidence effect of the detection efficiency for gamma rays which are emitted through cascade transition was appropriately simulated.

\section{Sampling and measurement}
The sampling of airborne radioactivity was started on 18 March, 2011, 
seven days after the Great East Japan Earthquake.
The sampling site was placed at the top of the building of the University of 
Tokushima, placed at 134$^{\circ}$33'E longitude, 34$^{\circ}$4'N latitude and 
15m altitude, about 700km southwest of Fukushima.
The airborne radioactive isotopes were collected by a high volume air sampler 
HVC-1000N provided by SIBATA whose sampling rate was 1m$^{3}$/min.
The filter for sampling was a commercial glass filter GB-100R provided by ADVANTEC 
with the dimension of 203mm$\times$254mm.
The efficiency for retaining particles with a size of 0.3$\mu$m is 99.88\%.
The sampling was started at 12:00 and continued 23 hours.

\begin{figure}[ht]
\begin{center}
\includegraphics[width=10cm]{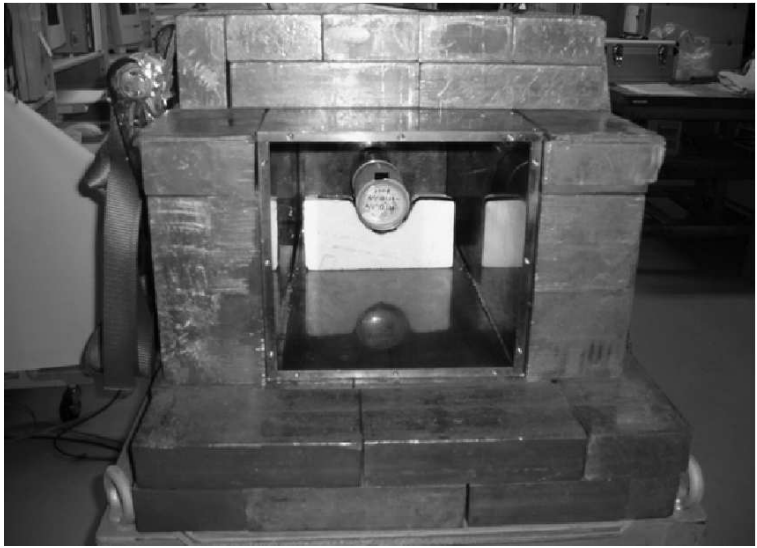}
\end{center}
\caption{
Ge detector system. The front shield was opened in this picture.
A sample container was put on a foam polystyrene mount.
}
\label{fg:Ge}
\end{figure}

The filter was striped into 1cm width bands and contained into a sample container
made of polycarbonate.
A sample container was placed in front of the end cap of a HPGe detector.
The distance between the end cap and the sample container was 0.3cm.
Fig.\ref{fg:Ge} shows the Ge detector system, whose shield was opened.
The Ge detector and the sample container was covered with 1cm thick OFHC
(Oxygen Free High Conductive) copper plates and 10cm thick lead bricks.
The total gamma ray background was reduced three orders of magnitude 
by the shield.
The signal from pre-amplifier was shaped by the shaping amplifier 
ORTEC 571.
The pulse height was digitized by  a multichannel analyzer.
The energy spectrum was stored into a hard disk every two hours and the data taking was 
continued for 24 hours after the end of sampling.
Since about 12 hours from the end of sampling, the background events is dominated by the ones from the progeny of 
Rn,
the data for the present work were taken between 16 hours and 24 hours after the 
end of sampling.

The Rn density was used for the check of the sampling efficiency of aerosol.
The annual modulation of Rn density in Tokushima has been measured for 17 years.
The density of Rn does not change every year in the same season\cite{Rn}.

\begin{figure}[ht]
\begin{center}
\includegraphics[width=10cm]{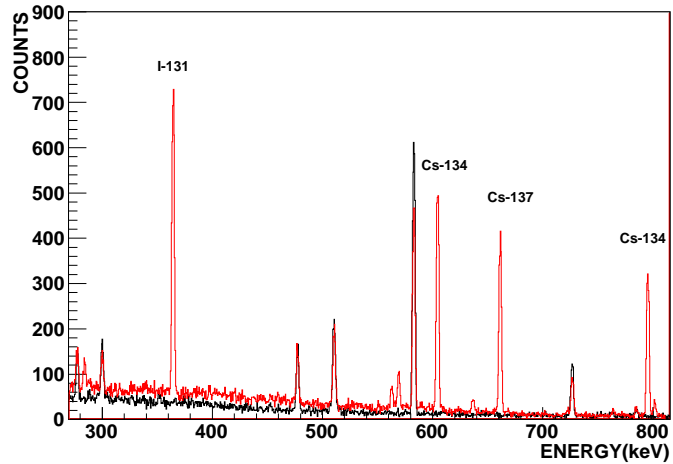}
\end{center}
\caption{
The energy spectra of gamma rays which was emitted by radioactive isotopes attached to 
a filter.
The red line is the energy spectrum taken on 6 April 2011.
The black line is the one taken before the Great East Japan Earthquake.
}
\label{fg:spect}
\end{figure}

The significant fission products and activation products were measured from 23 March, 2011.
The energy spectrum was shown in Fig.\ref{fg:spect}.
After 23 March, clear peaks due to $^{131}$I, $^{134}$Cs and $^{137}$Cs were
observed.

\section{Monte Carlo simulation to determine the efficiency}
The detection efficiency for gamma rays from $^{134}$Cs and $^{131}$I must be carefully estimated.
The detection efficiency is distorted by coincidence 
of some gamma rays which are emitted by cascade decay.
For example, the 604keV gamma ray from the excited state of $^{134}$Ba
is accompanied by the other gamma rays.
Consequently, the detection efficiency is distorted by the coincidence with other gamma rays.
The efficiency distortion depends on the geometrical distribution of the source and 
the detector.
However, the correction of the distortion is rather difficult because the geometrical 
arrangement may change by each measurement.

The Monte Carlo simulation is the good tool to determine the detection efficiency 
for a complex geometrical arrangement.
In the present work, Geant4.9.4.p02 was used to determine the efficiency.
Geant4 \cite{Geant4} is the simulation tool kit to simulate the transportation of 
gamma ray, beta ray and other ionizing radiation particles.
The class ``G4RadioactiveDecay'' in Geant4.9.4.p02 generates the radioactive decays 
of almost all the nuclei.
The properties of unstable nuclei, half life, decay mode, excited state, branching ratio
are listed in the class.
The cascade emission of gamma rays is properly simulated by the G4RadioactiveDecay 
class.

\begin{figure}[ht]
\begin{center}
\includegraphics[width=12cm]{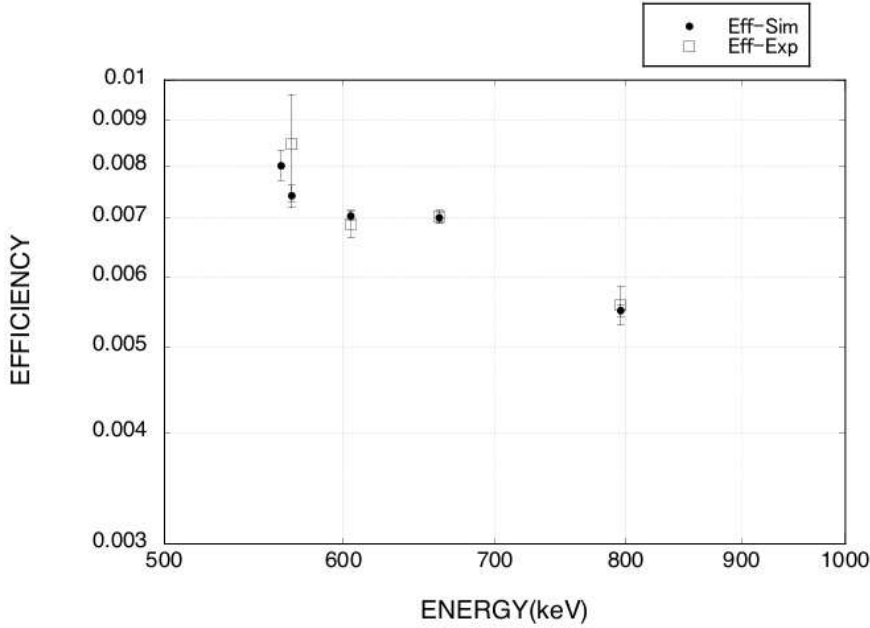}
\end{center}
\caption{
The detection efficiency of gamma rays. Box: Experimental result.
Circle: Simulated result.
Experimental result for 569keV and 563keV could not dissolved so that the efficiency was 
calculated as 569keV gamma ray.
}
\label{fg:sim}
\end{figure}

To verify the simulation, a simulation and a practical measurement were performed.
The measurement was performed by using the IAEA-444 standard source \cite{IAEA444}.
The standard source was contained in a U-8 pack, whose dimension was 
50.4mm$\phi\times$60.2mm.
The radioactive sources were uniformly mixed into soil.
The U-8 pack was put in front of the end cap of a HPGe detector.
The simulation  was performed with the same dimension.
The energy dependence of the detection efficiency which were derived by experiment and simulation was 
plotted in Fig.\ref{fg:sim}.
The Monte Carlo simulation well agreed with the experimental result.
Especially, the detection efficiency of 604keV gamma ray was lower than the energy dependence 
of efficiency which was fitted by polynomial function.
We determined the detection efficiency for the various shapes of sample containers by 
Monte Carlo simulation.

\section{Analysis}
\subsection{Correction of radioactive decay}
From the peak yields measured by Ge detector, the density of radioactivity
was calculated.
The effect of the radioactive decay is significant for calculating the proper 
radioactivity of short lived nuclei, for example, $^{131}$I.\@ 
The half life of  $^{131}$I is so short as 8.04 day that the decay between 
the sampling was not negligible.
The amount of attached nuclei on a filter $A$ is corrected by means of 
following procedure\cite{JAEA2010-039}.

First, the total number of attached nuclei $A_{T}$ is expressed as
\begin{equation}
A_{T}=\rho VT_{S}g,
\end{equation}
where $\rho$, $V$, $T_{S}$ and $g$ are the density of the nucleus in air, 
the volume of sampled air per unit time, the sampling time and the 
collection efficiency of filter, respectively.
During the sampling, the number of nuclei on the filter $A(t)$ increases 
as $P=\rho Vg$.
The number of the nuclei decreases by radioactive decay, thus
the equation is
\begin{equation}
\frac{dA(t)}{dt}=P-\lambda A(t).
\end{equation}
Solving the equation, one gets the actual number of nuclei attached on the filter 
at the end of sampling, say $A_{T}$ as,
\begin{equation}
A_{T}=A(T_{S})\frac{\lambda T_{S}}{1-\exp(-\lambda T_{S})},
\end{equation}
where, $\lambda=\ln 2/T_{1/2}$ is decay constant and $T_{1/2}$ is half-life.

After sampling fished, the nuclei decays with the decay constant $\lambda$, 
thus the number of the nuclei when the beginning of measurement is
\begin{equation}
A(T_{C})=A(T_{S})\exp(-\lambda T_{C}),
\end{equation}
where $T_{C}$ is the time interval between the ending of sampling and 
the starting of measurement.
 
The radioactive decay should be considered if the half-life is short.
The number of decay $\Delta N$ between the measurement time $T_{M}$ is 
expressed as 
\begin{eqnarray}
\Delta N & = & N(T_{C})-N(T_{C}+T_{M})  \nonumber \\
               & = & N(T_{C})\left\{ 1-\exp(-\lambda T_{M}) \right\}.
\end{eqnarray}

The present measurement, the peak yield $Y$ acquired by Ge detector is 
expressed as , 
\begin{equation}
Y=\epsilon a \Delta N,
\end{equation}
where $\epsilon$ and $a$ are the detection efficiency of gamma ray and the 
emitting ratio of the gamma ray.

The important parameters for the present measurement are listed in 
Table \ref{tb:param}.
\begin{table}[ht]
\caption{
The parameters of the present measurement.
}
\label{tb:param}
\begin{center}
\begin{tabular}{l|r} \hline 
Parameters  & Value \\ \hline
Sampling time $T_{S}$[hour] & 23 \\
Time interval $T_{C}$[hour] & 16 \\
Measurement time $T_{M}$[hour] & 8 \\
Decay constant of $^{131}$I $\lambda$[hour$^{-1}$]& $3.592\times 10^{-3}$ \\ \hline
\end{tabular}
\end{center}
\end{table}

\subsection{Efficiency of filter}
The commercial filter GB-100R provided by ADVANTEC was used for sampling 
the airborne radioactivity.
The filter efficiency of aerosol whose diameter is larger than 300nm is 
as large as 0.9999 \cite{ADVANTEC}.
The capture efficiency of airborne radioactive isotope depends on the chemical 
property.

\subsubsection{Cesium}
$^{137}$Cs and $^{134}$Cs are exhausted from a reactor attached on aerosol.
The size of aerosol which attaches cesium was measured as 
830$\sim$860nm \cite{Reineking}.
The filter efficiency for cesium was confirmed by measuring the density of $^{7}$Be.
The size of aerosol containing $^{7}$Be is the same as the one containing cesium.
The density of $^{7}$Be in Tokushima has been measured since 2005.
The density of $^{7}$Be in Tokushima was 2$\sim$6mBq/m$^{3}$,  
which agrees the results measured at another sites \cite{JER-2006}.
The filter efficiency of GB-100R for cesium isotopes is considered as $g=0.999$.

\subsubsection{Iodine}
The filter efficiency $g$ for iodine depends on the chemical structure of iodine.
The gaseous iodine such as I$_{2}$, HOI and CH$_{3}$I cannot be collected by 
normal glass filter.
Only a small fraction of iodine ion and chemical compounds of iodine which are 
attached on aerosol is cached by our glass filter.
The size of aerosol particle which iodine is attached is rather smaller than 
the size of $^{137}$Cs, the average size is 590$\sim$613nm \cite{Reineking}.
The filter efficiency for iodine attached on aerosol is treated as 0.999, however, 
the fraction of the particle iodine must be considered.

The fraction of iodine forms was investigated by Noguchi \cite{Noguchi}.
The detail of the fraction is shown in technical note on the monitoring 
of airborne radioactivity\cite{MEXT}.
The average fraction of iodine attached on aerosol is $0.20\pm0.1$. 
The total filter efficiency is the product of 0.999 and the average fraction of 
particle iodine, resulting $g=0.2\pm0.1$.

\section{Results and Discussion}
\begin{figure}[htb]
\begin{center}
\includegraphics[width=12cm]{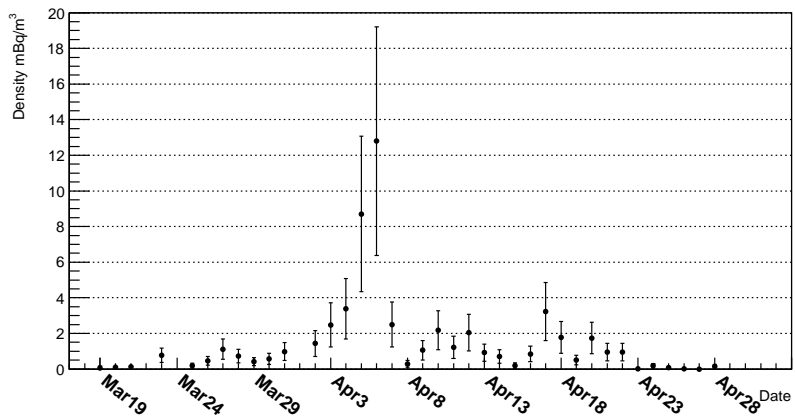}
\end{center}
\caption{
The daily variation in the density of  $^{131}$I in air.
}
\label{fg:density}
\end{figure}

The daily variation of $^{131}$I densities is
 shown in Fig. \ref{fg:density}.
The error of each data is dominated by the error of correction efficiency $g$.
The significant radioactivity was observed after 23 March.
The behavior of cesium isotopes was the same as iodine, however,  the concentration of cesium isotopes
 was one order smaller than the one of iodine.

\begin{figure}[htb]
\begin{center}
\includegraphics[width=12cm]{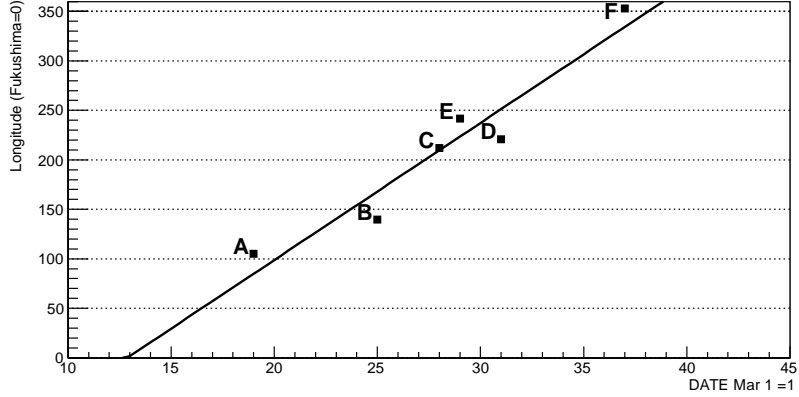}
\end{center}
\caption{
The dates which the maximum density of radioactivity was observed.
A: Seattle (USA)\cite{Diaz} \ B:Chapel Hill (USA)\cite{MacMullin} \ 
C:Huebla (Spain)\cite{Lozano} \ D:Orsay (France)\cite{IRSN} \ 
E:Thessaloniki (Greece)\cite{Monolo} \ F:Tokushima (Japan, This work)
}
\label{fg:longitude}
\end{figure}

The maximum of the density was observed on 6 April, about three weeks after the large vent.
This behavior cannot be explained that the radioactive plume came directly from 
Fukushima Daiichi Nuclear Plant.
The dates which the maximum density was measured in other sites have a strong 
correlation. 
The speed of the transportation of radioactive plume is explained that 
westerly wind brought the radioactive plume.
Fig.\ref{fg:longitude} shows the relationship between the dates which the  maximum 
radioactivity was observed in each cites.
The dates of each cites and their longitude have a strong correlation.
The speed of the plume was calculated from the linear fitting.
The speed was 40km/sec which agreed the speed of westerly wind.

In western side of American Continent (Seattle), the first significant observation was 
on 17 March and reached the maximum on 19 March\cite{Diaz}.
In eastern side of American Continent (Chapel Hill), 
the first observation was on 18 March,
however, the maximum was observed on 29 March. 
This discrepancy came from the rainfall on 20 March \cite{MacMullin}.
They observed three peaks of density. 
The first peak of density was observed on 25 March.
After the first peak, two peaks were observed on 30 March and 2 April.
The density of the first peak was reduced by rainfall, so we considered that the plume arrived on 
25 March.

The radioactive plume was brought to Europe by westerly wind.
In western Spain (Huevla), the maximum was reported on 28 March\cite{Lozano}.
However, the sampling was made only a few times, 15-17, 21-23 and 28-29 March.
After 28-29, they continuously  measured till 15-17 April.
In France (Orsay), the continuous measurement was reported by IRSN (Institut de 
Radioprotection et de S\^uret\'e Nucl\'eaire)\cite{IRSN}.
The first significant observation was on 25 March and the maximum 
was observed on 31 March.
In Greece (Thessaloniki), the first significant observation was on 26 March and 
the maximum was observed on 29\cite{Monolo}.
The relationship between the dates of maximum density and the longitude 
was well fitted by linear function.

To confirm the hypothesis that the radioactive plume which arrived on 6 April went around 
the Northern Hemisphere, further analysis was carry out.
The isotopic component in a plume from nuclear plant was dominated by 
cesium so that the ratio $R\equiv^{131}\mbox{I}/^{137}\mbox{Cs}$ is larger than 10
\cite{MEXT}.
The ratio becomes smaller during the long travel by the decay of $^{131}$I and 
by dissolving into rainwater.
The ratio in Seattle was reported as large as 31,  while, the ratios in Europe 
was between 10 to 4.
The decrease in $R$ is more rapid than the radioactive decay of the isotopes.

The effect on the isotopic ratio was clearly observed by Asian sites.
In Taiwan, the value of $R$ was dropped from $\sim$1 to $\sim 0.3$ between the end of
March and the beginning of April\cite{Taiwan}.
In Vietnam, the value $R$ decreased exponentially\cite{Vietnam} and the ratio was small.

\begin{figure}[htb]
\begin{center}
\includegraphics[width=12cm]{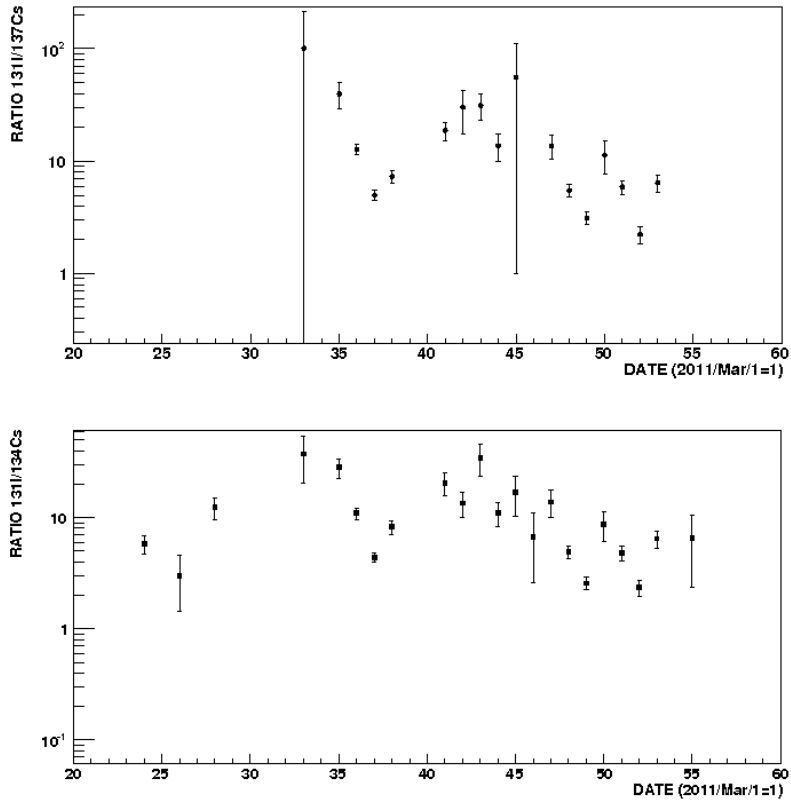}
\end{center}
\caption{
The daily variation in the ratio of $R\equiv^{131}\mbox{I}/^{137}\mbox{Cs}$
(Upper) and $R\equiv^{131}\mbox{I}/^{134}\mbox{Cs}$ (Lower).
}
\label{fg:ratio}
\end{figure}

The temporary decrease of $R$ was clearly observed in Tokushima.
The daily variation in the ratio $R$ in Tokushima was analyzed as shown in 
Fig.\ref{fg:ratio}.
In the beginning of April (1st April $\sim$ 2nd April), the value of $R$ was large, 
which suggests the radioactive plume came directly from Fukushima.
During 3 to 7 in April, the ratio $R$ temporarily decreased as $R\simeq4$.
From the isotopic component, the biggest peak around 6 April was 
caused by the radioactive plume exhausted on 12$\sim$15 in March and the 
plume traveled around the Northern Hemisphere.
In Vietnam, the altitude is so small that the radioactive plume did not pass through there.

The concentration of measured radioactivity was about five orders of magnitude
smaller than the regulation in Japan.
The estimated dose was negligibly low expecting no health effect in 
western Japan.

\section{Acknowledgements}
The authors thank the University of Tokushima for supporting the 
continuous measurement.

\end{document}